\documentclass[twocolumn,showpacs,preprintnumbers,pra]{revtex4}
\usepackage{graphicx}
\usepackage{dcolumn}
\usepackage{bm}
\usepackage{color}

\begin{document}

\title{Ultrabright Backward-wave Biphoton Source}
\author{Chih-Sung Chuu}
\email{cschuu@stanford.edu}
\author{S. E. Harris}
\affiliation{Edward L. Ginzton Laboratory, Stanford University, Stanford, California
94305, USA}

\begin{abstract}
We calculate the properties of a biphoton source based on resonant backward-wave spontaneous parametric down-conversion. We show that the biphotons are generated in a single longitudinal mode having a subnatural linewidth and a Glauber correlation time exceeding 65 ns.
\end{abstract}

\maketitle

The parametric interaction of electromagnetic waves where the signal and idler propagate in opposite directions was first suggested  by Harris \cite{Harris66} and has now been extensively studied \cite{Chemla74,Yariv89,Ding96,Su06}. Its special feature is, that because of the internal feedback provided by the backward wave, the interaction becomes temporally unstable at a finite crystal length. For many decades the experimental challenge has been the lack of an appropriate nonlinear material for phase matching. The breakthrough came recently, where by extending the techniques of quasiphase matching \cite{Fejer92} to sub-micron periodicity \cite{Canalias03,Canalias05}, Canalias and Pasiskevicius have demonstrated the first mirrorless optical parametric oscillator \cite{Canalias07}.

In this paper we suggest and calculate the properties of parametric down converter where because the parametric interaction is of the backward-wave type, the linewidth for parametric gain and spontaneous emission is about 40 times narrower than for a forward wave interaction. By placing the nonlinear crystal within a resonant cavity, the counterpropagating signal and idler photons are generated in a single longitudinal mode with a linewidth that is less than that of typical radiative transitions, and have a Glauber correlation time  greater than 65 ns.

Biphoton sources play a central role in applications of quantum information processing such as linear optical quantum computation (LOQC) \cite{Knill01} and long distance quantum communication \cite{Briegel98}. Biphotons of subnatural linewidth and long correlation times are particularly desirable for these applications; in part, this is because the subnatural linewidth allows photon entanglement to be stored in atomic ensemble
 memories  \cite{Memory}. This is essential for efficient generation of multiphoton entanglement in LOQC \cite{Browne05,Bodiya06} as well for  applications involving quantum repeaters \cite{Duan01}. The long correlation time also allows interference of independent photon sources, a key element for producing multiphoton entanglement without the need for synchronization~\cite{Halder07}.

The most widely used source of biphotons is forward wave spontaneous parametric down-conversion (SPDC) in nonlinear crystals, wherein a pump photon splits into two co-propagating photons of lower frequencies. However, the loose constraint of phase matching results in linewidths that are typically on the order of THz and are too broad to efficiently interact with atoms. Passive filtering with narrowband filters can be employed to reduce the linewidth but, at the same time, decrease the biphoton generation rate. Forward wave SPDC with active filtering has been demonstrated by resonating the signal and idler fields with an optical cavity \cite{Cavity}. But, because of the broad gain linewidth,  multiple cavity modes are resonated simultaneously. Additional spectral filtering, such as an etalon locked to the resonant cavity is thus necessary for obtaining a single-mode output. Narrowband biphotons may also be generated in cold atoms by using the techniques of cavity quantum electrodynamics \cite{Keller04} or of electromagnetically induced transparency \cite{Balic05}.

\begin{figure}
\centering
\includegraphics[width=9cm]{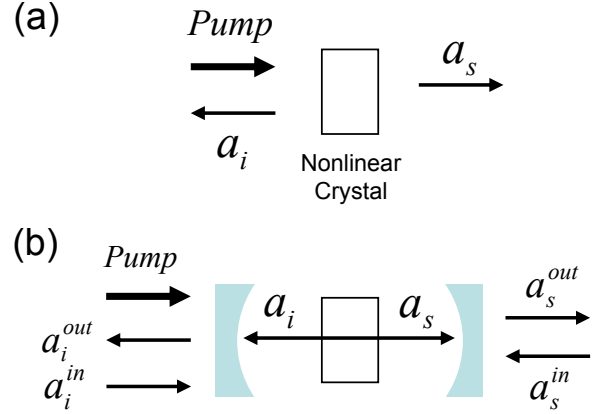}
\caption{\label{fig1} (a) Schematic of backward-wave spontaneous parametric down-conversion (SPDC). The operators $a_s$ and $a_i$ denote the signal and idler fields. (b) Schematic of backward-wave SPDC within a resonant cavity. The operators $a^{\rm in}_{s,i}$ and $a^{\rm out}_{s,i}$ denote the incident fields and output fields, respectively.}
\end{figure}

We develop the theory in the Heisenberg picture. We assume the pump is a monochromatic classical field at frequency $\omega_p$ and take the signal and idler frequencies as $\omega_s=\omega$ and $\omega_i=\omega_p-\omega$. In the absence of a cavity [Fig.~\ref{fig1}(a)], the output of the backward-wave SPDC may be described by the frequency domain operators $a_s(\omega,z)=b_s(\omega,z) \exp[ik_s(\omega)z]$ and $a_i(\omega_i,z)=b_i(\omega_i,z) \exp[ik_i(\omega_i)z]$, where the operators $b_s(\omega,z)$ and $b_i(\omega_i,z)$ vary slowly with distance $z$. The coupled equations for $b_s(\omega,z)$ and $b^{\dagger}_i(\omega_i,z)$ are
\begin{eqnarray}
\frac{\partial b_s (\omega,z)}{\partial z} &=& i \kappa b^{\dagger}_i (\omega_i,z) \exp[i \Delta k(\omega) z] \nonumber \\
\frac{\partial b_i^{\dagger} (\omega_i,z)}{\partial z} &=& i \kappa b_s (\omega,z) \exp[- i \Delta k(\omega) z],
\label{Coupled_Eq}
\end{eqnarray}
where $\kappa$ is the coupling constant. With $L$ denoting the crystal length, the quantities $a_s(\omega,L)$ and $a^{\dagger}_i(\omega_i,0)$ may then be expressed in terms of the vacuum fields at the input of the crystal, $a_s(\omega,0)$ and $a^{\dagger}_i(\omega_i,L)$,
\begin{eqnarray}
a_s(\omega,L) &=& A(\omega) a_s(\omega,0) + B(\omega) a^{\dagger}_i(\omega_i,L)\nonumber \\
a^{\dagger}_i(\omega_i,0) &=& C(\omega) a_s(\omega,0) + D(\omega) a^{\dagger}_i(\omega_i,L).
\label{backward_coupled_eq}
\end{eqnarray}

The spectral power density \cite{Harris07} at the signal frequency is $S(\omega) = \int_{- \infty}^{\infty} \langle a^{\dagger}_s (\omega) a_s (\omega') \rangle \exp[i(\omega - \omega') t] d \omega'$. Noting the commutators 
$[a_j(\omega_1,0),a^{\dagger}_k(\omega_2,0)]=[a_j(\omega_1,L),a^{\dagger}_k(\omega_2,L)]=\frac{1}{2\pi}\delta_{jk}\delta(\omega_1-\omega_2)$,
\begin{equation}
S(\omega) = \frac{1}{2 \pi}\left| B(\omega) \right|^2.
\label{B2}
\end{equation}

If the gain is small, the coefficients in Eq.~(\ref{backward_coupled_eq}) are given by $A(\omega)=\exp[i k_s(\omega) L]$, $D(\omega)=\exp[i k_i(\omega_i)L]$, $C(\omega)=~B^*(\omega) \exp \{ i[k_s(\omega)+k_i(\omega_i)]L \}$, and 
\begin{eqnarray}
B(\omega) &=& i \kappa L\ {\rm sinc} \left[ \frac{\Delta k(\omega) L}{2} \right] \nonumber \\
&& \times \exp \left\{i \left[ \frac{\Delta k(\omega)}{2} + k_s(\omega) + k_i(\omega_i) \right] L \right\}.
\end{eqnarray}
The spectral power density at the signal frequency is then $S(\omega) = \frac{1}{2 \pi}\kappa^2 L^2 {\rm sinc}^2 [\Delta k(\omega) L/2]$, where the \textit{k}-vector mismatch $\Delta k (\omega) \approx (v^{-1}_{s}+v^{-1}_{i}) \Delta \omega_s$ with $v_{s} = \partial \omega / \partial k_s(\omega)$ and $v_{i} = \partial \omega_i / \partial k_i(\omega_i)$ denoting the group velocities at the signal and idler frequencies, and $\Delta \omega_s$ equal to the detuning of the signal frequency from line center. The gain linewidth for a backward-wave interaction is thus $\Delta \omega_G \approx 1.77 \pi /[(v^{-1}_{s}+v^{-1}_{i})L]$. 

\begin{figure}
\centering
\includegraphics[width=9cm]{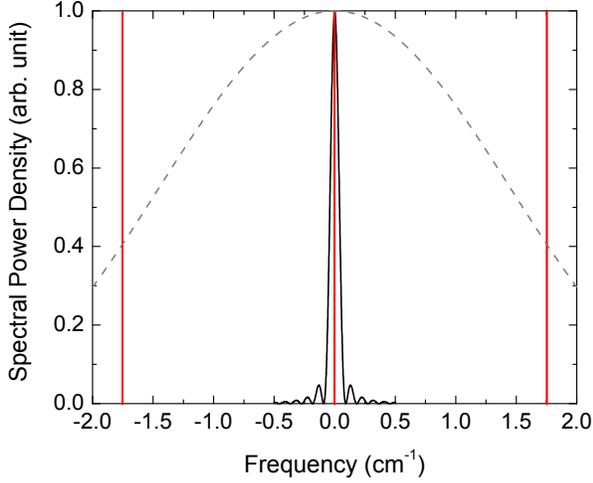}
\caption{\label{fig2} The solid (dashed) curve represents the spectral power density of the backward- (forward-) wave spontaneous parametric down-conversion at the signal frequency. The vertical (red) lines denote the signal frequencies of adjacent mode pairs of a resonant cavity that are separated by the cluster spacing $\Delta \Omega_{\rm Cl}$. The central vertical line is taken at the degenerate frequency. }
\end{figure}

If we assume a 3-cm long periodically-poled potassium titanyl phosphate (KTP) crystal pumped by 532 nm laser and quasiphase-matched at the degenerate frequency, then $\Delta \omega_G \cong 2 \pi \times 0.08$ cm$^{-1}$ or $2 \pi \times 2.4$ GHz. As compared to a forward wave interaction in which $\Delta k (\omega) \approx (v^{-1}_{s}-v^{-1}_{i}) \Delta \omega_s$, the gain linewidth for a backward-wave interaction in a crystal of equal length is reduced by a factor of $(v^{-1}_{s}+v^{-1}_{i})/|v^{-1}_{s}-v^{-1}_{i}| \cong 38$ (Fig.~\ref{fig2}). 

We next consider the case where a nonlinear generating crystal of length $L$ is placed inside a resonant cavity of the same length [Fig.~\ref{fig1}(b)]. We assume that only a single pair of signal and idler fields coincide with the $q$th and $r$th cavity modes. (This single-mode assumption will be justified below). For this cavity case it is convenient to describe the quantum fields by  time domain operators $a_s(t,z)=b_s(t) \exp(-i \Omega_q t) \sin(q \pi z/L)$ and $a_i(t,z)=b_i(t) \exp(-i \Omega_r t) \sin(r \pi z/L)$, where $b_s(t)$ and $b_i(t)$ are the fields internal to the cavity and vary slowly with time. $\Omega_q$ and $\Omega_r$ are the cold cavity frequencies. The coupled equations for the slowly varying operators are
\begin{eqnarray}
\frac{\partial b_s (t)}{\partial t} + \frac{\Gamma_s}{2} b_s (t) &=& - i \kappa_1\ b^{\dagger}_i(t) + \sqrt{\gamma_s}\ b^{\rm in}_s(t) \nonumber \\
\frac{\partial b^{\dagger}_i (t)}{\partial t} + \frac{\Gamma_i}{2} b^{\dagger}_i (t) &=& i \kappa_1\ b_s(t) + \sqrt{\gamma_i}\ b^{\rm in \dagger}_i(t), \label{eq:coupled1} 
\end{eqnarray}
where the envelope quantities $ b^{\rm in}_s(t)$ and $b^{\rm in \dagger}_i(t)$ are the fields incident on the resonant cavity, and $\Gamma_s$ and $\Gamma_i$ are the total cavity decay rates. With $E_p$ as  the electric field strength of the pump, and with a \textit{k}-vector mismatch $\Delta k' = k_p - q \pi/L - r \pi/L$, the coupling constant $\kappa_1=\frac{1}{2} d \epsilon_0^2 c^2 \eta_s \eta_i (\omega_s \omega_i)^{1/2} E_p \exp ( i \Delta k' L/2 ) {\rm sinc} ( \Delta k' L/2 )$. With $\gamma_s$ and $\gamma_i$ denoting the output coupling rates, the slowly varying output fields $b^{\rm out}_s(t)$ and $b^{\rm out \dagger}_i(t) $ are
\begin{eqnarray}
b^{\rm out}_s(t) &=& \sqrt{\gamma_s}\ b_s(t) - b^{\rm in}_s(t) \nonumber \\
b^{\rm out \dagger}_i(t) &=& \sqrt{\gamma_i}\ b^{\dagger}_i(t) - b^{\rm in \dagger}_i(t). 
\label{eq:coupled2}
\end{eqnarray}

We solve for the output fields by transforming the coupled equations to the frequency domain with the Fourier pair $b(t) = \int_{- \infty}^{\infty} b(\omega') \exp (-i \omega' t) d \omega'$ and $b(\omega') = \frac{1}{2 \pi} \int_{- \infty}^{\infty} b(t) \exp (i \omega' t) d t$. The slowly varying quantities are then converted to fast varying analytic signals (nonzero for positive frequencies) by $a(\omega_{s,i})=b(\omega_{s,i}-\Omega_{q,r})$ and $a^{\dagger}(\omega_{s,i})=b^{\dagger}(\omega_{s,i}+\Omega_{q,r})$. The output fields $a^{\rm out}_s (\omega)$ and $a^{\rm out \dagger}_i (-\omega_i)$ may be written in terms of the incident fields $a^{\rm in}_s (\omega)$ and $a^{\rm in \dagger}_i (-\omega_i)$,
\begin{eqnarray}
a^{\rm out}_s (\omega) &=& A_1(\omega)\ a^{\rm in}_s (\omega) + B_1(\omega)\ a^{\rm in \dagger}_i (- \omega_i) \nonumber \\
a^{\rm out \dagger}_i (- \omega_i) &=& C_1(\omega)\ a^{\rm in}_s (\omega) + D_1(\omega)\ a^{\rm in \dagger}_i (- \omega_i), \label{eq:ABCD1}
\end{eqnarray}
where for small gain the coefficients are
\begin{eqnarray}
A_1(\omega) &=& \frac{\gamma_s - \Gamma_s/2+i (\omega-\Omega_q)}{\Gamma_s/2-i (\omega-\Omega_q)} \nonumber\\
B_1(\omega) &=& \frac{-i \kappa_1 \sqrt{\gamma_s \gamma_i}}{[\Gamma_s/2-i (\omega-\Omega_q)][\Gamma_i/2+i (\omega_i-\Omega_r)]} \nonumber \\
C_1(\omega) &=& \frac{i \kappa_1 \sqrt{\gamma_s \gamma_i}}{[\Gamma_s/2-i (\omega-\Omega_q)][\Gamma_i/2+i (\omega_i-\Omega_r)]} \nonumber \\
D_1(\omega) &=& \frac{\gamma_i - \Gamma_i/2-i (\omega_i-\Omega_r)}{\Gamma_i/2+i (\omega_i-\Omega_r)} \label{eq:ABCD2}
\end{eqnarray}
and, for a lossless cavity are related by unitary conditions.

We use Eq.~(\ref{eq:ABCD1}) and (\ref{eq:ABCD2}) to derive the spectral and temporal properties of the biphotons. Noting the commutator, $[a^{\rm in}_j(\omega_1),a^{\rm in \dagger}_k(\omega_2)]=[a^{\rm in}_j(\omega_1),a^{\rm in}_k(\omega_2)]=\frac{1}{2\pi}\delta_{jk}\delta(\omega_1-\omega_2)$, the spectral power density at the signal frequency [Eq.~(\ref{B2})] is
\begin{equation}
S_1(\omega) = \frac{8 \gamma_s \gamma_i \kappa^2_1}{\pi [4(\omega-\Omega_q)^2 + \Gamma_s^2][4(\omega_i-\Omega_r)^2 + \Gamma_i^2]}.
\label{eq:spectrum}
\end{equation}
For exact phase matching ($\Delta k' =0$), the biphoton linewidth is $\Delta \omega = [(\sqrt{\Gamma_s^4+6 \Gamma^2_s \Gamma^2_i + \Gamma_i^4}-\Gamma^2_s-\Gamma^2_i)/2]^{1/2}$, and the total paired count rate is
\begin{equation}
R_1 = \frac{1}{2 \pi} \int_{-\infty}^{\infty} |B_1(\omega')|^2 d \omega' = \frac{4 \gamma_s \gamma_i \kappa^2_1}{\Gamma_s \Gamma_i (\Gamma_s+\Gamma_i)}.
\end{equation}

The Glauber two-photon correlation function is $G^{(2)}(t_s,t_i) = \langle a^{\rm out \dagger}_i (t_i) a^{\rm out \dagger}_s (t_s) a^{\rm out}_s (t_s) a^{\rm out}_i (t_i) \rangle$, where $t_s$ and $t_i$ are the arrival times of the signal and idler photons, respectively. Defining the time delay $\tau=t_i-t_s$, the time domain Glauber correlation function may be written as~\cite{Harris07}
\begin{eqnarray}
G^{(2)}(\tau)= \left| \frac{1}{2 \pi}\int_{-\infty}^{\infty} A_1(\omega')C_1^*(\omega') e^{i \omega' \tau} {\rm d}\omega' \right|^2+ \nonumber \\
 \left| \frac{1}{2 \pi} \int_{-\infty}^{\infty} |B_1(\omega')|^2 d \omega' \right|^2.
\label{eq:fullG2}
\end{eqnarray}
The second term in Eq.~(\ref{eq:fullG2}) is independent of $\tau$ and results from accidental two photon events. To the extent that the generation rate of biphotons is small as compared to the inverse of the temporal length of the biphoton, this term may be neglected. The first term in Eq.~(\ref{eq:fullG2}) then evaluates to 
\begin{eqnarray}
G^{(2)}(\tau)= \frac{4 \Gamma_s \Gamma_i \kappa^2_1}{(\Gamma_s + \Gamma_i)^2} \times
\left\{ \begin{array}{ll}
 e^{\Gamma_s \tau} &\mbox{, $\tau<0$ } \\
 e^{-\Gamma_i \tau} &\mbox{, $\tau>0$ }.
       \end{array} \right. 
\label{eq:G2}
\end{eqnarray}
The asymmetry in $\tau$ in Eq.~(\ref{eq:G2}) is due to the order of detection of the signal and idler photons. The photon arriving at the detectors earlier in time triggers the correlation measurement. The shape of $G^{(2)}(\tau)$ is thus determined by the photon arriving later in time. For example, when the idler photon arrives first so that $t_s > t_i$ and $\tau < 0$, $G^{(2)}(\tau)$ is determined by the cavity decay rate of the signal photons. The Glauber correlation time (full width at half maximum) is then $T_c = (\ln 2) (1/\Gamma_s+1/\Gamma_i)$ and the coherence time \cite{Pomarico09} is $1/\Gamma_s+1/\Gamma_i$. With $\Delta_{s,i}$ and $r_{s,i}$ denoting the spacing of the cavity modes and the mirror reflectivity, respectively, the output coupling rate is $\gamma_{s,i} = \Delta_{s,i} (1-r_{s,i})$. With $\xi_{s,i}$ defined as the single-pass power loss in the crystal, the total cavity decay rates are $\Gamma_{s,i}=2 \xi_{s,i} \Delta_{s,i} + \gamma_{s,i}$.

\begin{figure}
\centering
\includegraphics[width=9.5cm]{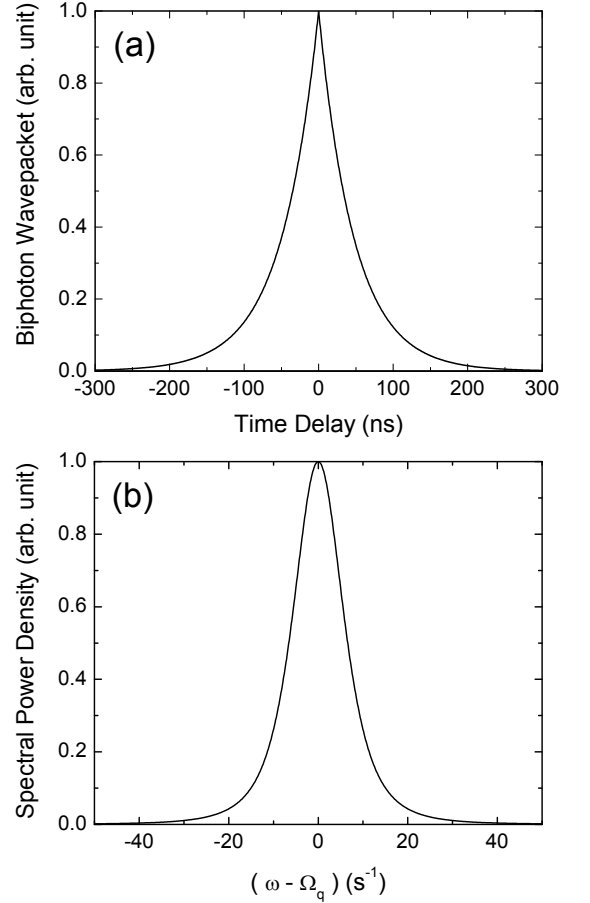}
\caption{\label{fig3}(a) Time domain biphoton wavepacket of the backward-wave spontaneous down-conversion within a resonant cavity. Though difficult to see, this curve is slightly asymmetric [see Eq.~(\ref{eq:G2})]. (b) Spectral power density at the signal frequency.}
\end{figure}

To justify the validity of the single-mode assumption, we compare the cluster spacing $\Delta \Omega_{\rm Cl}$, i.e. the frequency separation between two pairs of signal and idler cavity modes satisfying energy conservation, to the gain linewidth $\Delta \omega_G$ of the backward-wave SPDC. The cluster spacing can be obtained by solving $M(\omega) \Delta \Omega^2_{\rm Cl}+N(\omega) \Delta \Omega_{\rm Cl} = \pm 1$ with $M(\omega)=[L/(2 \pi c)]\{2[n^{'}_s (\omega)+n^{'}_i(\omega_i)]+\omega_s n^{''}_s(\omega)+\omega_i n^{''}_i(\omega_i)\}$ and $N(\omega)=[L/(\pi c)][n_s-n_i+\omega_s n^{'}_s (\omega_s)-\omega_i n^{'}_i (\omega_i)]$, where $n_s$ and $n_i$ are the refractive indices at the signal and idler frequencies, and $n^{'}_{s,i}$ and $n^{''}_{s,i}$ are the first and second frequency derivatives, respectively \cite{Eckardt91}. 

As an example we consider a resonated 3-cm long periodically poled KTP crystal. We take the resonant cavity to be the same length as the nonlinear crystal and to have a finesse of 1000.  Then $\Delta \Omega_{\rm Cl} \cong 2 \pi \times 1.75$ cm$^{-1}$ and $\Delta \omega_G \cong 2 \pi \times 0.08$ cm$^{-1}$. Since $\Delta \Omega_{\rm Cl} > \Delta \omega_G$, when the cavity is appropriately tuned, there will be only a single mode-pair within the backward-wave gain linewidth (Fig.~\ref{fig2}). 

We estimate the linewidth, Glauber correlation time, and spectral brightness of the biphotons based on the above parameters. To ensure high purity of biphoton generation, we assume a pump power of 770 $\mu$W which is far below the threshold power so that the generation rate is small as compared to the inverse of the coherence time. For the optimum case of exact phase matching ($\Delta k' =0$), the total paired count rate is $R_1 \cong 1.31 \times~10^5\ {\rm s}^{-1}$ and the biphoton linewidth $\Delta \omega\cong 2 \pi \times 2.1$~MHz is smaller than that of typical atomic transitions. The spectral brightness is then $R_1/\Delta \omega \cong 6.25 \times 10^4$~s$^{-1}$MHz$^{-1}$ or $8.16 \times 10^4$~s$^{-1}$MHz$^{-1}$ per mW of pump power. The time domain biphoton wavepacket is given by Eq.~(\ref{eq:G2}) and has a sharp fall-off [Fig.~\ref{fig3}(a)] due to its Lorentzian-shaped spectrum [Eq.~(\ref{eq:spectrum}) and Fig.~\ref{fig3}(b)]. The Glauber correlation time $T_c\cong68$~ns is approximately the sum of the ring-down times at the signal and idler frequencies.

To construct a source as described above we expect to use a KTP crystal which is periodically poled with a periodicity of $\Lambda=872$ nm. A 532 nm laser may be used as the pump source to generate signal and idler photons at the degenerate frequency of 1.064 $\mu$m. The pump and signal will be polarized along the crystal y axis, the idler is polarized along the crystal z axis, and quasiphase matching is accomplished in third order ($m=3$) so that $k_p = K_G + k_s - k_i$, where the lattice \textit{k}-vector $K_G=2 \pi m / \Lambda$. The ratio of the spectral brightness of this resonant backward-wave source, as compared to a non-resonant forward wave source of the same material, pumping power, and length is about 80000.

In conclusion we have described a narrowband biphoton source that utilizes resonant backward-wave parametric down-conversion. The narrow backward-wave gain linewidth allows a single mode output with both greatly increased spectral brightness, and also, the generation of biphotons that are sufficiently long that they may be amplitude or phase modulated by high speed light modulators \cite{Specht09,Belthangady09}.  If successfully constructed, applications may include quantum communication, quantum memories, and enhanced resistance against narrowband interference for quantum key distribution \cite{Belthangady10}.

This work was supported by the U.S. Air Force Office of Scientific Research and the U.S. Army Research Office. C.-S.C. acknowledges support from the Taiwan National Science Council (NSC98-2917-I-564-136).

\bibliography{backward}
% Produces the bibliography via BibTeX.

\end{document}